\documentclass{ptephy}

\usepackage{color,ulem}

\begin{document}

\title{Production of hyperon resonances induced by kaon on a deuteron target}

\author{\name{J. Yamagata-Sekihara}{1,\ast}, \name{T. Sekihara}{1}, and \name{D. Jido}{2,3}}

\address{\affil{1}{Institute of Particle and
  Nuclear Studies, High Energy Accelerator Research Organization
  (KEK), 1-1, Oho, Ibaraki 305-0801, Japan}
\affil{2}{Yukawa Institute for Theoretical Physics, 
Kyoto University, Kyoto 606-8502, Japan}
\affil{3}{J-PARC Branch, KEK Theory Center,
  Institute of Particle and Nuclear Studies, 
  High Energy Accelerator Research Organization (KEK),
  203-1, Shirakata, Tokai, Ibaraki, 319-1106, Japan}
\email{yamajun@post.kek.jp}}

\begin{abstract}%
The $K^- d \to \pi YN$reaction is theoretically studied so
as to investigate the $K^{-}$-induced production of the
hyperon resonances $\Sigma (1385)$ and $\Lambda (1405)$.
For this purpose we take into account the $p$-wave amplitudes
for the meson-baryon two-body scatterings as well as the
$s$-wave amplitudes.  Due to the fact that the hyperon
resonances are selectively produced from the $\bar{K} N$
channel in this reaction, the $\Lambda (1405)$ peak appears
at 1420 MeV, which implies that $\Lambda (1405)$ and
$\Sigma (1385)$ could be separately seen in the missing
mass spectrum of the emitted nucleon in the $K^{-} d \to
n X$ reaction.  The $\pi Y$ invariant mass spectrum in this
study is consistent with experimental data both for $\Sigma
(1385)$ and $\Lambda (1405)$.  The pion exchange
contributions are also included and are found to give smooth
background without distorting the peak structure of the hyperon
resonances.
\end{abstract}


\maketitle
\section{Introduction}
\label{sec:1}

Properties of hyperon resonances are important to understand role of
strangeness in hadron physics.  Especially, understanding of hyperon
resonances located below the $\bar K N$ threshold, such as
$\Lambda(1405)$ and $\Sigma(1385)$, is essential for physics of
strangeness in nuclear matter and nuclei.  In addition, because
baryonic resonances decay into mesons and a baryon in strong
interactions, one can learn basic interactions between hadrons with
strangeness from the properties of hyperon resonances.

Among the various hyperon resonances, recent attention is particularly
focused on the $\Lambda(1405)$ hyperon resonance with $J^{P}=1/2^{-}$
and $I=0$. The $\Lambda(1405)$ resonance has been considered for long
time as a quasibound state of $\bar KN$~\cite{Dalitz:1959dn}, being
extremely important resonance to understand $\bar KN$ interaction.
Theoretically, $\Lambda(1405)$ is successfully reproduced as
dynamically generated states in coupled-channels approach based on
chiral dynamics~\cite{Kaiser:1995eg,Oset:1997it,Oller:2000fj,Lutz:2001yb,Oset:2001cn,Hyodo:2002pk,Jido:2003cb,Borasoy:2005ie,Borasoy:2006sr,Hyodo:2011ur,Ikeda:2011pi,Ikeda:2012au},
and this approach has confirmed that the $\Lambda(1405)$ is
predominantly described by meson-baryon
components~\cite{Hyodo:2008xr}.  Recently it has been also pointed out
in Ref.~\cite{Jido:2003cb} that the $\Lambda(1405)$ is composed by two
resonance states having different coupling nature and that the one
which dominantly couples to the $\bar KN$ channel is located at 1420
MeV instead of nominal 1405 MeV. Since the resonance position of the
$\Lambda(1405)$ in the $\bar KN$ channel is strongly related to the
strength of the $\bar KN$ interaction, it is very important to observe
$\Lambda(1405)$ spectra in the $\bar KN \to \pi\Sigma$ channel and pin
down the resonance position of the $\Lambda(1405)$ seen below the
$\bar KN$ threshold.

Several ideas to observe the $\Lambda(1405)$ initiated by the $\bar
KN$ channel have been proposed in Refs.~\cite{Jido:2003cb,Hyodo:2004vt,Lutz:2004sg,magas,Geng:2007hz,Jido:2009jf,Jido:2010rx,Jido:2012cy,Esmaili:2009rf}.  
One of the promising ways to form the
$\Lambda(1405)$ in the $\bar KN$ channel is to use nuclear reactions
with $K^{-}$ beam, such as in-flight $K^{-}d \to \Lambda(1405)n$, as
discussed in Refs.~\cite{Jido:2009jf,Jido:2010rx}, in which Fermi
motion of nucleon and $\bar K$ multi-scattering with nucleons help to
create $\Lambda(1405)$ by $\bar KN$ below its threshold.
In Ref.~\cite{Jido:2009jf} it has been found that, in the $K^{-}d \to
\Lambda(1405)n$ reaction with intermediate $K^{-}$ beam energy,
$\Lambda(1405)$ production dominantly takes place in double scattering
process, in which the incoming $K^{-}$ kicks one of the nucleons in
deuteron in forward direction and looses its energy suitably to form
$\Lambda(1405)$ with the other nucleon.  Although a single scattering
process also contributes to the $\Lambda(1405)$ production, for
energetic incoming $K^{-}$ with several hundreds MeV$/c$ in the
laboratory frame the contribution is small due to insufficient Fermi
motion of nucleon in deuteron.
In addition, thanks to the fact that strangeness is brought into the system 
by the incoming $K^{-}$, $\Lambda(1405)$ is formed selectively by the $\bar KN$ 
channel\footnote{Formation of $\Lambda(1405)$ by the $\pi \Sigma$ channel
would take place by multi-pion exchange between two baryons. However, multi-pion 
exchange processes are absolutely negligible for the resonance formation 
with in-flight $K^{-}$.}. This is a good advantage over $\Lambda(1405)$ 
productions induced by a nonstrange particle, in which strangeness has to 
be created in the reaction process and can be carried by the baryon producing 
$\Lambda(1405)$.
In fact, there has been already an old bubble-chember experiment
observing $\Lambda(1405)$ in $K^{-}d \to \pi^{+} \Sigma^{-} n$
at $K^{-}$ momenta between 686 and 844 MeV/c~\cite{Braun:1977wd}.
Although the statistics was not so high, the experiment found clearly that
the $\Lambda(1405)$ spectrum has a peak at 1420 MeV. 
Forthcoming experiments in J-PARC with high-intensity kaon beam are expected
for more detailed and precise information on the properties of 
$\Lambda(1405)$~\cite{Noumi:JPARC}. Recently it has been 
pointed out in Ref.~\cite{Miyagawa:2012xz} that the way of the determination 
of the kinematics in the loop given in Ref.~\cite{Jido:2009jf}
would break the three-body unitarity. Later in Ref.~\cite{Jido:2012cy}
the authors have proposed several options for the kaon energy in the loop and found
that the threshold 
effect does not affect the spectrum shape. In addition, since
the $\Sigma(1385)$ resonance is far from the $\bar KN$ threshold, 
this issue is irrelevant for the $\Sigma(1385)$ resonance. 

In this paper, we apply this $K^{-}$-induced production off deuteron
target also to the $\Sigma(1385)$ resonance, which has $J^{P}=3/2^{+}$
and $I=1$ and is also located below the $\bar KN$ threshold. The
$\Sigma(1385)$ is understood well in quark model point of view; it is
classified in flavor decuplet.  Nevertheless, there are few
theoretical discussion on $\Sigma(1385)$ formation in the $\bar KN$
channel. In addition, for precise determination of the $\Lambda(1405)$
properties from experiments, it is very important to understand
contributions of $\Sigma(1385)$ in the $\Lambda(1405)$ production
process, because the $\Lambda(1405)$ and $\Sigma(1385)$ resonances are
located at very similar energy having tens MeV widths and their
spectra overlap each other.
This reaction gives also important informations on the subthreshold productions of
$\Lambda(1405)$ and $\Sigma(1385)$, which will be significant in the strange few-body
bound systems of $\bar KNN$ \cite{Yamazaki:2002uh}, $\bar K KN$ \cite{Jido:2008kp} and $\pi \Lambda N$ \cite{Gal:2008eq}.
We also evaluate background contributions
against $K^{-}d \to Y^{*}N$ by considering pion exchange contributions, in
order to make further comparison with the experimental spectra
observed in Ref.~\cite{Braun:1977wd}.

This paper is organized as follows.  In Sec.~\ref{sec:2} we explain
our scheme to calculate the production cross sections for the $\Sigma
(1385)$ as well as $\Lambda (1405)$.  In Sec.~\ref{sec:3} we show our
numecrical results of the calculations for the production cross
sections.  We also discuss nonresonant background contributions rather
than $\Sigma (1385)$ and $\Lambda (1405)$ to the mass spectra in this
section.  Section~\ref{sec:4} is devoted to the summary of this study.

\section{Formulation}
\label{sec:2}

\begin{figure}[t]
\begin{center}
\includegraphics[width=6.0cm]{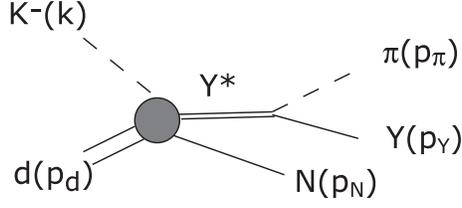}
\caption{\label{fig:1}
Kinematics for the $K^-d\to \pi Y N$ reaction.
}
\end{center}
\end{figure}

In this study we consider the $K^-d\to \pi^0\Lambda
n,~(\pi\Sigma)^0n,~\pi^-\Lambda p$, and $(\pi\Sigma)^-p$ reactions so
as to discuss the production of the hyperon resonances $\Sigma(1385)$
and $\Lambda(1405)$ initiated by the $\bar{K} N$ channel.  The
kinematical variables are given in Fig.~\ref{fig:1}. 
We discuss the kinematical aspects
of the reactions in Sec.~\ref{sec:2A}, and in the proceeding sections 
we discuss the dynamics of the reaction including the scattering amplitudes for
$K^{-}d$ reaction in Sec.~\ref{sec:2B} and the meson-baryon scattering
amplitudes in Appendix.

\subsection{Reaction kinematics\label{sec:2A}}

The reaction $K^-d\to \pi YN$ requires five variables to completely fix
the phase-space of the three-body final state~\cite{Nakamura:2010zzi}.
In this study we are interested in the mass spectra of the $\pi Y$
systems, thus we choose as the five variables the $\pi Y$ invariant mass
$M_{\pi Y}$, the solid angle of the final $N$ in the total
center-of-mass frame $\Omega_N$, and the solid angle of the final
$\pi$ in the $\pi Y$ center-of-mass $\Omega^*_\pi$.  Then, the cross
section of the reaction is calculated by,
\begin{equation}
d\sigma
=\left(\frac{M_dM_YM_N}{(2\pi)^4 4k_{\rm c.m.}E_{\rm c.m.}}\right)
|{\mathcal T}|^2
|\mbox{\boldmath $p$}_N||\mbox{\boldmath $p$}^*_\pi|
dM_{\pi Y} d\cos\theta_Nd\Omega^*_\pi~,
\label{eq:1}
\end{equation}
where $M_{d}$, $M_{Y}$, and $M_{N}$ are initial-state deuteron and
final-state hyperon and nucleon masses in the reaction, respectively,
$E_{\rm c.m.}$ is the total center-of-mass energy, $k_{\rm c.m.}$ is
the $K^{-}$ momentum in the total center-of-mass frame, $\theta_N$ is
the scattering angle of $N$ with respect to $K^{-}$ in the total
center-of-mass frame, and $\mathcal T$ is the $T$-matrix of the reaction.
In this form the azimuthal angle of the $\Omega_N$ is integrated.

Now let us pin down the reaction mechanism.  Since we want to produce
the hyperon resonances in the final state, the diagrams in which $\pi$
and $Y$ come from the same vertices are essential.  Here,
following Ref.~\cite{Jido:2009jf} we evaluate the impulse and
double-scattering amplitudes for the productions of the hyperon
resonances.

\begin{figure}[t]
\begin{center}
\includegraphics[width=8.1cm]{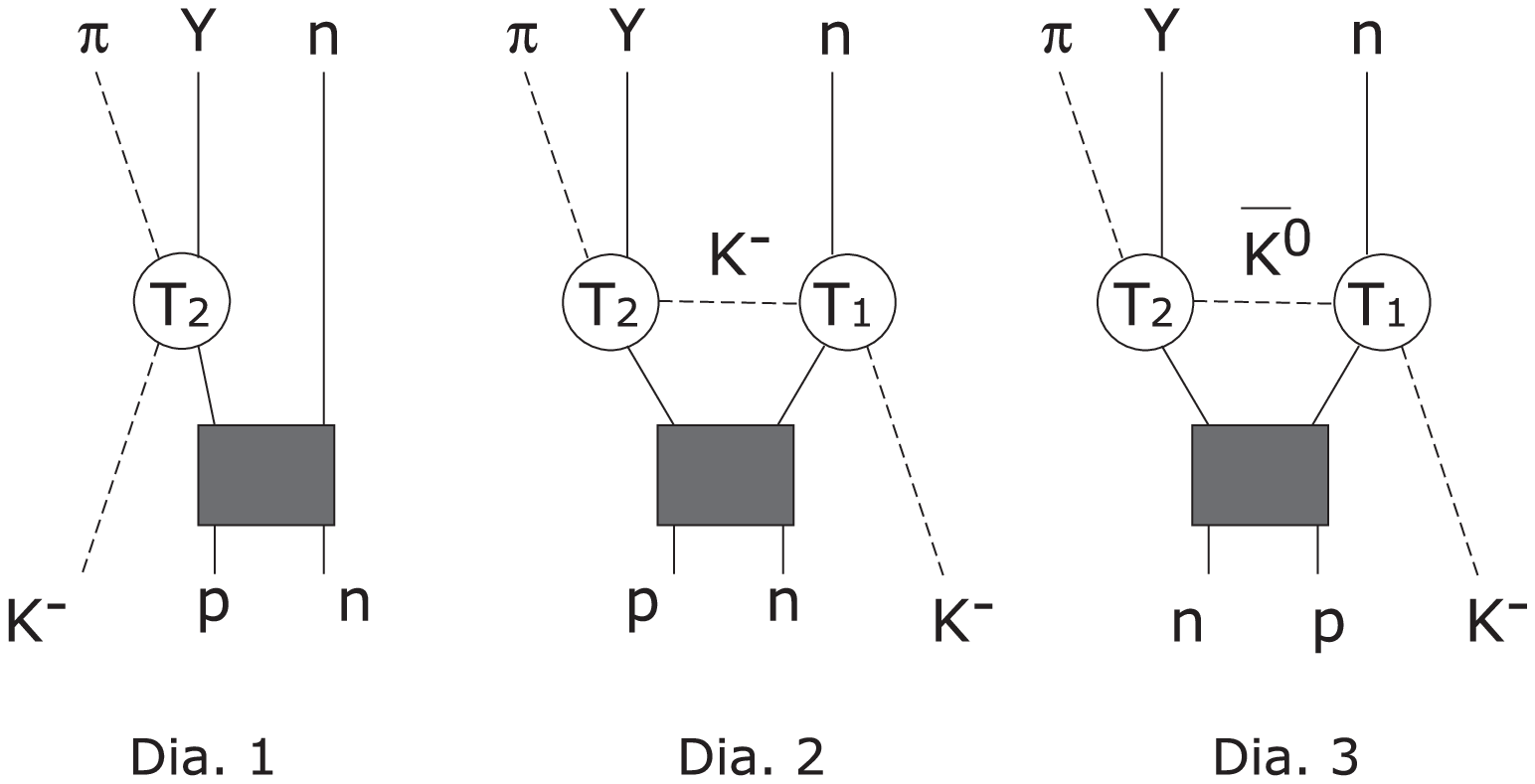}
\caption{\label{fig:2} Diagrams for the $K^-d\to \pi Y n$ reaction.
  In the diagrams $T_1$ and $T_2$ denote the scattering amplitudes for
  $K^{-} N\to{\bar K}n$ and ${\bar K}N\to \pi Y$, respectively.  }
\end{center}
\end{figure}

First we consider the $K^-d$ reaction with neutron
emission in the final state.  In this reaction the relevant diagrams are given in
Fig.~\ref{fig:2}.  Diagram 1 in Fig.~\ref{fig:2} corresponds to the
impulse process for the hyperon resonance production, whereas diagrams
2 and 3 are double scattering processes.  We emphasize that the
conservation of the strangeness in the strong interaction guarantees
the production of hyperon resonances from the ${\bar K}N$ channel both in
the impulse and the double scattering processes.  The $\pi Y$
systems are fixed as charge zero ones,
$\pi^0\Lambda$,~$\pi^+\Sigma^-$, $\pi^-\Sigma^+$, and $\pi^0\Sigma^0$.
The $\pi^0\Lambda$ ($\pi^0\Sigma^0$) mass spectrum will be dominated
by the $\Sigma(1385)$ ($\Lambda (1405)$), since it is purely $I=1$
($0$) channel.  The $\pi^+\Sigma^-$ and $\pi^-\Sigma^+$ mass spectra
can contain $\Sigma(1385)$ in addition to $\Lambda(1405)$, since
$\Sigma(1385)$ has the branching ratio $12\%$ to the $\pi\Sigma$
channel.

\begin{figure}[t]
\begin{center}
  \includegraphics[width=5.4cm]{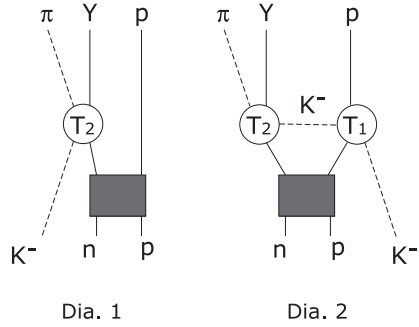}
  \caption{\label{fig:2.5} Diagrams for the $K^-d\to \pi Y p$
    reaction.  In the diagrams $T_1$ and $T_2$ denote the scattering
    amplitudes for $K^{-}p\to K^{-}p$ and $K^{-}n\to \pi Y$,
    respectively.  }
\end{center}
\end{figure}

Second we consider the $K^-d$ reaction with proton
emission in the final state.  The relevant diagrams are given in Fig.~\ref{fig:2.5}.  In
this case we have only two diagrams so as to emit the final-state
$\pi$ and $Y$ from same vertex; one is the impulse diagram and the
other the double scattering diagram.  The $\pi Y$ systems are
$\pi^-\Lambda$, $\pi^-\Sigma^0$, and $\pi^0\Sigma^-$, and all of them
contain only the $\Sigma^-(1385)$ contribution as hyperon resonances
of interest.

Now that we have determined the reaction diagrams, all we have to do
is the evaluation of the scattering amplitudes, which contains
dynamics of the productions of the hyperon resonances.

\subsection{$K^{-} d$ Scattering amplitudes\label{sec:2B}}

In this section we formulate the scattering amplitudes of the $K^-d$
reaction given in Figs.~\ref{fig:2} and \ref{fig:2.5}.  These
amplitudes are composed of ${\bar K}N\to{\bar K}N$ and $\pi Y$
amplitudes, kaon propagator, and deuteron wave function.  We study both $\Sigma(1385)$ and $\Lambda(1405)$ in the reaction by including 
$s$- and $p$-wave contributions in the ${\bar K}N$ amplitudes.  It is important that the $p$-wave
contribution requires the nucleon spin component in the deuteron wave
function, which is not needed in case of the $\Lambda(1405)$
production with the $s$-wave ${\bar K}N$
amplitudes~\cite{Jido:2009jf}.

First we formulate the amplitude of the impulse approximation given in
Figs.~\ref{fig:2}(1) and \ref{fig:2.5}(1).  According
to Ref.~\cite{Jido:2009jf} and taking into account the spin component
inside the deuteron, we can write the scattering amplitude for the
impulse approximation as,
\begin{align}
& {\mathcal T}^a_{\pi Y n(1)}=T_{K^-p\to \pi Y}(M_{\pi  Y},\hat{p_\pi}\cdot\hat{k})
(S^\dagger)^a \tilde{\varphi}(|\mbox{\boldmath
 $p$}_n-\frac{\mbox{\boldmath $p$}_d}{2}|)~~,\\ 
& {\mathcal T}^a_{\pi Y p(1)}=T_{K^-n\to \pi Y}(M_{\pi  Y},\hat{p_\pi}\cdot\hat{k}) 
(S^\dagger)^a \tilde{\varphi}(|\mbox{\boldmath $p$}_p-\frac{\mbox{\boldmath
    $p$}_d}{2}|)~~.
\end{align}
Here $T_{\bar{K}N\to MB}$ is the $\bar{K}N\to MB$ amplitude in the $2\times 2$
matrix form, which compensates the Pauli spinor for baryons, and is
function of center-of-mass energy $M_{\pi Y}$ and the angle
$\hat{k}\cdot\hat{p}_\pi$ in meson-baryon center-of-mass frame,
$(S^\dagger)^a=-i\sigma^2\sigma^a/\sqrt{2}$~$(a=1,~2,~3)$ the spinor
component for each nucleon inside the deuteron in $2\times 2$ matrix
form, and $\tilde{\varphi}(p)$ is the momentum representation of the
deuteron wave function with momentum $p$, for which we neglect the
$d$-wave component and use a parameterization of the $s$-wave
component given by an analytic function~\cite{Lacombe:1981eg} as,
\begin{eqnarray}
\tilde{\varphi}(p) = 
\sum_{j=1}^{11} \frac{C_j}{p^2+m_j} , 
\end{eqnarray}
with $C_{j}$ and $m_{j}$ determined in Ref.~\cite{Machleidt:2000ge}.

Second let us consider the double scattering amplitudes.  According
to Ref.~\cite{Jido:2009jf} and taking into account the spin component
inside the deuteron, the double scattering amplitudes are formulated
as,
\begin{align} 
{\mathcal T}^a_{\pi Y n (2)}=&\int\frac{d^3q}{(2\pi)^3}
  T_{K^-p\to\pi Y}(M_{\pi Y}, \hat{q}\cdot\hat{p_\pi})(S^\dagger)^a
  T^t_{K^-n\to K^-n}(W,\hat{k}\cdot\hat{q})
  \nonumber\\
  &\times\frac{\tilde\varphi(|\mbox{\boldmath $q$}+\mbox{\boldmath
      $p$}_n-\mbox{\boldmath $k$}-\mbox{\boldmath
      $p$}_d/2|)}{q^2-m^2_{K^-}+i\epsilon} ~~,
  \\
  {\mathcal T}^a_{\pi Y n (3)}=&-\int\frac{d^3q}{(2\pi)^3} T_{{\bar
      K}^0n\to \pi Y}(M_{\pi Y},\hat{q}\cdot{\hat p_\pi})(S^\dagger)^a
  T^t_{K^-p\to{\bar K}^0n}(W,{\hat k}\cdot{\hat q})
  \nonumber\\
  &\times\frac{\tilde\varphi(|\mbox{\boldmath $q$}+\mbox{\boldmath
      $p$}_n-\mbox{\boldmath $k$}-\mbox{\boldmath
      $p$}_d/2|)}{q^2-m^2_{\bar K^0}+i\epsilon} ~~,
  \\
  {\mathcal T}^a_{\pi Y p(2)}=&\int\frac{d^3q}{(2\pi)^3} T_{K^-n\to \pi
    Y}(M_{\pi Y},{\hat q}\cdot{\hat p_\pi})(S^\dagger)^a T^t_{K^-p\to
    K^-p}(W,{\hat k}\cdot{\hat q})
  \nonumber\\
  &\times\frac{\tilde\varphi(|\mbox{\boldmath $q$}+\mbox{\boldmath
      $p$}_p-\mbox{\boldmath $k$}-\mbox{\boldmath
      $p$}_d/2|)}{q^2-m^2_{K^-}+i\epsilon} ~~,
\end{align}
where $q^0$ and $W$ is fixed as, $q^0=k^0+M_1-p^0_N$ and
$W=\sqrt{(M_1+k^0)^2-\mbox{\boldmath $k$}^2}$, with $M_1$ the
first-scattered nucleon mass and $p^0_N$ the energy of the final-state
nucleon.  This prescription of $q^{0}$ takes account of the $NN$ potential 
in deuteron nonperturbatively as suggested in the Watson 
formulation~\cite{Jido:2012cy}.
The superscript $t$ denotes the transposition of the matrix
in the spin space.

In our approach, the ${\bar K}N\to{\bar K}N$ and $\pi Y$ amplitudes are essential to the production of the hyperon resonances.
For the meson-baryon amplitudes we apply the so-called chiral unitary
approach~\cite{Kaiser:1995eg,Oset:1997it,Oller:2000fj,Oset:2001cn,
  Lutz:2001yb,Hyodo:2002pk,Jido:2003cb}.  In chiral unitary approach,
$\Lambda(1405)$ is dynamically generated by the unitarized
coupled-channel method based on chiral dynamics without explicit
poles~\cite{Hyodo:2008xr}. In contrast, $\Sigma(1385)$ can be
included as an explicit pole in the $p$-wave kernel interactions in
the coupled-channels~\cite{Jido:2002zk}.
The details of the description of the ${\bar K}N\to{\bar K}N$ and $\pi Y$ amplitudes for the $\Lambda(1405)$ and $\Sigma(1385)$ are given in Appendix~\ref{appendix:A}.

The chiral unitary amplitude has been calculated in the center of mass frame of the two-body meson baryon system. Since the deuteron wavefunction has been calculated in the rest frame of the deuteron, we calculate the $K^{-}d \to \pi YN$ amplitude in the lab frame, in which the target deuteron is at rest. Thus, we make a transformation of the amplitude obtained in the two-body center of mass frame to the baryon rest frame using the method shown in Appendix~\ref{AppB}. For the $p$-wave amplitude, we define the off-shell behavior by using a form factor 
\begin{equation}
     f_{\Lambda}(|\vec q\,|) = \frac{\Lambda^{2}}{\Lambda^{2} + \vec q^{\, 2}}
\end{equation}
with $\Lambda = 630$ MeV, which has been used in the chiral unitary model for the $s$-wave. 

The total amplitude for $K^-d\to\pi Y n$ and $\pi Y p$ is
given by the coherent sum of the impulse and double scattering
contributions as,
\begin{align}
& {\mathcal T}^a_{\pi Y n}={\mathcal T}^a_{\pi Y n(1)}+{\mathcal T}^a_{\pi Y n(2)}+{\mathcal T}^a_{\pi Y n(3)} ~, \\
& {\mathcal T}^a_{\pi Y p}={\mathcal T}^a_{\pi Y p(1)}+{\mathcal T}^a_{\pi Y p(2)} ~, 
\end{align}
respectively.  Then, squared amplitude with spin-summed for final
state and averaged in the initial state (deuteron) is given as,
\begin{equation}
|{\mathcal T}|^2=\frac{1}{3}\sum_{a=1}^{3} 
\textrm{tr}[{\mathcal T}^a({\mathcal T}^a)^\dagger]~~,
\end{equation}
with taking the trace of the $2 \times 2$ spin space matrix by
``$\textrm{tr}$".


%

\subsection{Estimation of the pion exchange contributions\label{sec:2D}}
\begin{figure}[t]
\begin{center}
\includegraphics[width=2.7cm]{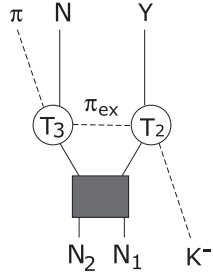}
\caption{\label{fig:14}
Diagrams of the $\pi$ meson exchange contribution.
$T_2$ and $T_3$ denote the scattering amplitudes for ${\bar K}N\to \pi Y$ and $\pi N\to\pi N$, respectively.
}
\end{center}
\end{figure}

It is instructive to estimate the pion exchange contribution to the production of the hyperon resonances, which mainly comes from the amplitudes without $\pi Y$ correlation in the final state.
In the estimation of the pion exchange amplitude, we evaluate the diagrams given in Fig.~\ref{fig:14}.
The particles, which are correlated to the pion exchange process, are listed in Table~\ref{tab:1}.
They contain the $\pi N\to\pi N$ amplitude, which do not appear in the usual diagrams for the production of hyperon resonances given in Figs.~\ref{fig:2} and \ref{fig:2.5}.
\begin{table}[t]
\begin{center}
\caption{\label{tab:1}Possible exchange pions for the diagram shown in Fig.~\ref{fig:14}}
\begin{tabular}{l|ccc}
\hline
\hline
Reaction&$N_1$&$N_2$&$\pi_{\rm ex}$\\
\hline
$K^-d\to \pi^+\Sigma^- n$&$p$&$n$&$\pi^+$\\
&$n$&$p$&$\pi^0$\\
$K^-d\to\pi^-\Sigma^+n$&$p$&$n$&$\pi^-$\\
$K^-d\to\pi^0\Sigma^0n$&$n$&$p$&$\pi^-$\\
&$p$&$n$&$\pi^0$\\
$K^-d\to\pi^0\Lambda n$&$n$&$p$&$\pi^-$\\
&$p$&$n$&$\pi^0$\\
$K^-d\to\pi^-\Lambda p$&$n$&$p$&$\pi^-$\\
&$p$&$n$&$\pi^0$\\
\hline
\hline
\end{tabular}
\end{center}
\end{table} 
For the $\pi N\to\pi N$ amplitude, we take an empirical amplitude~\cite{CNS}
up to $p$-wave including the $\Delta(1232)$ resonance, which are obtained based on the
observed scattering data. The off-shell extrapolation is done in the same way to the kaon exchange. 
We show the Dalitz plot in order to see the energy range of the $\pi N$ amplitude.
\begin{figure}[t]
\begin{center}
\includegraphics[width=8.0cm]{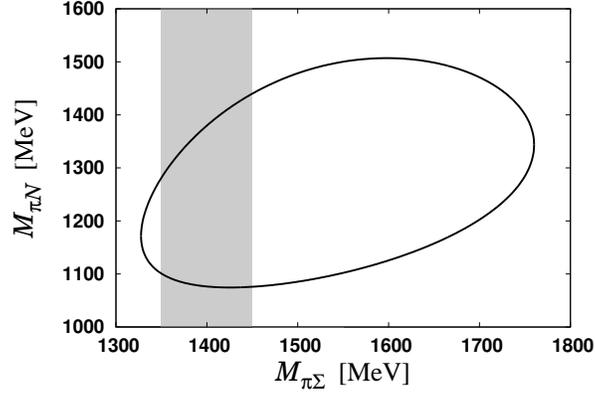}
\caption{\label{fig:dalitz} Dalitz plot for $\pi ^{0} \Sigma ^{0} n$
  final state with 800 MeV/c incident $K^-$ momenta.  Masked area
  shows the $\pi Y$ energy range ($1350$--$1450$ MeV) we considered in
  this work.  }
\end{center}
\end{figure}
In Fig.~\ref{fig:dalitz}, we have found the $\pi N$ energy range, which is considered from $\Lambda(1405)$ and $\Sigma(1385)$ productions, includes $\Delta(1232)$ resonance.
Therefore, the contribution of $\Delta(1232)$ production seems to be main smooth background from $\pi N$ scattering in $\Lambda(1405)$ and $\Sigma(1385)$ productions.

\section{Numerical results\label{sec:3}}
In this section we show our numerical results for the production of
hyperon resonances $\Sigma(1385)$ and $\Lambda(1405)$ in the $K^-d$
reactions.  Using Eq.~(\ref{eq:1}), we can evaluate the differential
cross section as,
\begin{equation}
\frac{d^2\sigma}{dM_{\pi Y}d\cos\theta_N}=\frac{M_dM_YM_N}{(2\pi)^44k_{\rm c.m.}E_{\rm c.m.}}|\mbox{\boldmath $p$}_N||\mbox{\boldmath $p$}^*_\pi|\int d\Omega^*_\pi|{\mathcal T}|^2~,
\end{equation}
with the $K^-d\to\pi YN$ scattering amplitude ${\mathcal T}$ which includes the pion exchange process.
Then, integrating $\cos\theta_N$, one can obtain the $\pi Y$ mass spectrum,
\begin{equation}
\frac{d\sigma}{dM_{\pi Y}}=\frac{M_dM_YM_N}{(2\pi)^44k_{\rm c.m.}E_{\rm c.m.}}|\mbox{\boldmath $p$}_N||\mbox{\boldmath $p$}^*_\pi| \int^1_{-1}d\cos\theta_N\int d\Omega^*_\pi|{\mathcal T}|^2
\label{eq:20}
\end{equation}
Further, integrating $M_{\pi Y}$ with appropriate range ($M_{\rm min}$, $M_{\rm max}$) for the hyperon resonances, the production cross section for the hyperon resonances is obtained as,
\begin{equation}
\sigma=\frac{M_dM_YM_N}{(2\pi)^44k_{\rm c.m.}E_{\rm c.m.}} \int^{M_{\rm max}}_{M_{\rm min}}dM_{\pi Y}|\mbox{\boldmath $p$}_N||\mbox{\boldmath $p$}^*_\pi|\int^1_{-1}d\cos\theta_N\int d\Omega^*_\pi|{\mathcal T}|^2~.
\end{equation}
It is also interesting to calculate the angular dependence of the production 
by, 
\begin{equation}
\frac{d \sigma}{d \cos \theta _{N}} = 
\frac{M_dM_YM_N}{(2\pi)^44k_{\rm c.m.}E_{\rm c.m.}} 
\int^{M_{\rm max}}_{M_{\rm min}}dM_{\pi Y} |\mbox{\boldmath $p$}_N| 
|\mbox{\boldmath $p$}^*_\pi| 
\int d\Omega^*_\pi|{\mathcal T}|^2~.
\label{eq:22}
\end{equation}

In this study we evaluate the $K^-d$ amplitudes in Figs.~\ref{fig:2}
and \ref{fig:2.5} by using chiral unitary approach for the description
of the meson-baryon scattering amplitudes.  In
this approach $\Lambda(1405)$ is dynamically generated without
introducing explicit poles in the $s$-wave, whereas $\Sigma(1385)$ is
included as an explicit pole in the $p$-wave amplitude.

First we compare our results with experiments in Sec.~\ref{sec:3A}.
Next in Sec.~\ref{sec:3B} theoretical studies of the production of the
hyperon resonances are given.

\subsection{Production of hyperon resonances -- comparison with
  experimental data\label{sec:3A}}

\subsubsection{$\Sigma(1385)$ production\label{sec:3A1}}

First of all, we show the results of the $\Sigma(1385)$ production and
make a comparison with the experimental data~\cite{Braun:1977wd}.  For
the $\Sigma(1385)$ production, it is better to see the
$K^-d\to\pi^-\Lambda p$ reaction, in which $\Lambda(1405)$ does not
contribute to the $\pi ^{-} \Lambda$ mass spectrum.
Here, we include the pion exchange contribution in addition to the kaon exchange contribution for the hyperon resonance production.

\begin{figure}[t]
\begin{center}
\includegraphics[width=8.0cm]{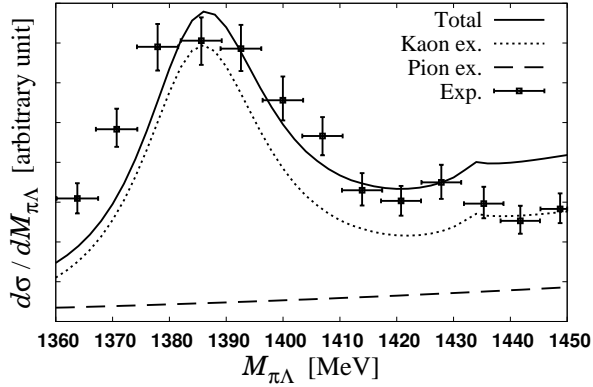}
\caption{\label{fig:piLambda1} $\pi^-\Lambda$ invariant-mass spectrum
  of the $K^{-} d \to \pi ^{-} \Lambda p$ reaction in arbitrary units
  at 800 MeV$/c$ incident $K^-$ momentum.  The solid line denotes the
  present calculation, and the dotted line and dashed line are the contribution from the kaon exchange and that from the pion exchange, respectively.
Experimental data are taken from
  Ref.~\cite{Braun:1977wd} at $K^{-}$ momenta between 686 MeV$/c$ and 
  844 MeV$/c$.  }
\end{center}
\end{figure}

Using Eq.~(\ref{eq:20}), we show the $\pi^-\Lambda$ mass spectrum in
the $K^-d\to\pi^-\Lambda p$ reaction in Fig.~\ref{fig:piLambda1},
together with the experimental data~\cite{Braun:1977wd}, which includes the
background contribution.  Here the
kaon momentum is fixed as 800 MeV$/c$ in our calculation, whereas the
range of the initiated kaon momentum is from 686 to 844 MeV$/c$ in the
experiment~\cite{Braun:1977wd}.  As seen in the figure, we can well reproduce
the experimental mass spectrum with the $\Sigma(1385)$ peak around
1385 MeV of the $\pi^-\Lambda$ invariant mass.

Next, let us evaluate the cross section of the $\Sigma(1385)$
production. In principle, such cross sections for the resonances can be evaluated
theoretically by taking the residue of the scattering amplitude at the resonance pole.
Here we take a more phenomenological way for the comparison with experimental data as follows.
Since the pion exchange contribution gives only the non-resonance background,
we take only the kaon exchange contribution for the production cross section,
and integrate it in the range of $M_{\rm min}=1370$ MeV and $M_{\rm max}=1400$ MeV:
\begin{equation}
\sigma_{\Sigma^*}=\frac{1}{0.88}\int^{M_{\rm max}}_{M_{\rm min}}dM_{\pi^-\Lambda}\frac{d\sigma_{\rm F.G}}{dM_{\pi^-\Lambda}}~,
\label{eq:23}
\end{equation}
where factor $1/0.88$ comes from the branching ratio of
$\Sigma(1385)\to\pi\Lambda$, $88\%$ and $\sigma_{\rm F.G.}$ is the cross section by foreground process (kaon exchange contribution).  We obtain the $\Sigma(1385)$
production cross section as 179 $\mu \text{b}$ with 800 MeV$/c$
incident $K^-$, whereas the experimental value observed in the
$K^-d\to\pi^-\Lambda p$ reaction is reported to be $252\pm 30$ $\mu
\text{b}$ at 778 MeV$/c$ of the incident $K^-$ momentum
\cite{Braun:1977wd}.  As one can see, our cross section is consistent
with the experimental value.  There is, however, small difference
between our theoretical production cross section and the experimental one.  
This difference may come from the ways to subtract background contribution 
of the resonance.
In Ref.~\cite{Braun:1977wd}, the background contributions have been 
estimated by fitting the mass spectrum in a sum of Legendre polynomials
together with a relativistic Brite-Wigner form for the resonance. 
In the  theoretical side, we have estimated the production cross section
by using the diagrams in which $\pi^{-}$ and $\Lambda$ are emitted 
from the same vertex and we have not subtracted
the non-resonant background appearing in the $K^-n \to \pi ^- \Lambda$ amplitude
against the $\Sigma(1385)$, which is clearly seen above the 1400 MeV
of the invariant mass $\pi^-\Lambda$.  We also note that the value of the
production cross section depends on the choice of the range of the invariant mass
integration.

At last we should note that above the 1420 MeV of the $\pi^-\Lambda$ invariant mass the mass spectrum is not suppressed although this is a far above the $\Sigma(1385)$ energy region.
This comes from $I=1$ non-resonant background in the ${\bar K}N$ scattering, 
which can interfere to $\Lambda(1405)$.

\subsubsection{$\Lambda(1405)$ production\label{sec:3A2}}
Next we show the results of the $\Lambda(1405)$ production in the $K^-d$ reaction.
For the $\Lambda(1405)$ production we choose $K^-d\to\pi^+\Sigma^- n$ reaction and calculate the $\pi^+\Sigma^-$ mass spectrum.
Here we note that $\pi^+\Sigma^-$ spectrum is contributed not only from $\Lambda(1405)$ but also from $\Sigma^0(1385)$, whose branching ratio to $\pi\Sigma$ is about 12$\%$.

\begin{figure}[t]
\begin{center}
\includegraphics[width=8.0cm]{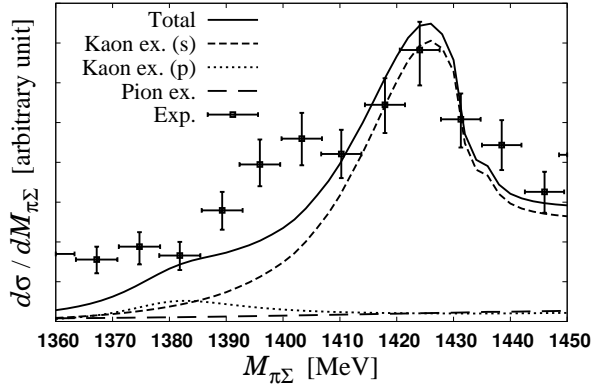}
\caption{\label{fig:pipSigm} $\pi^+\Sigma^-$ invariant-mass spectrum
  of the $K^{-} d \to \pi ^{+} \Sigma ^{-} n$ reaction in arbitrary
  units at 800 MeV$/c$ incident $K^-$ momentum. The solid line denotes
  the result of the full calculation with both $s$- and $p$-wave
  contributions, whereas the dashed (dotted) line denotes the
  contribution from the kaon exchange $s$- ($p$-) wave component in the ${\bar
    K}N\to \pi ^+ \Sigma ^-$ amplitude. The dahsed line corresponds to
  the result in Ref.~\cite{Jido:2009jf}.  
The long-dashed line denote the pion exchange contribution.
Experimental data are taken
  from Ref.~\cite{Braun:1977wd} at $K^{-}$ momenta between 686 MeV$/c$
  and 844 MeV$/c$.  }
\end{center}
\end{figure}

First, we show the $\pi^+\Sigma^-$ mass spectrum of the
$K^-d\to\pi^+\Sigma^- n$ reaction in Fig.~\ref{fig:pipSigm}, together
with the experimental data~\cite{Braun:1977wd}.
In the previous paper~\cite{Jido:2009jf}
the $K^-d$ reaction with only $s$-wave meson-baryon amplitude has been 
calculated, whereas in this study we include both $s$- and $p$-wave meson-baryon
amplitudes.  In Fig.~\ref{fig:pipSigm}, we plot the full
calculation with coherent sum of $s$- and $p$-wave contributions by
the solid line, whereas the $s$- ($p$-) wave contribution in ${\bar K}
N \to \pi ^+ \Sigma ^-$ amplitude ($T_2$ in Figs.~\ref{fig:2} and
\ref{fig:2.5}) is shown by the dashed (dotted) line. The $K^{-}N\to{\bar K}N$ amplitudes $T_{1}$ in
Figs.~\ref{fig:2} and \ref{fig:2.5} are fixed to be coherent sum of
the $s$- and $p$-wave contributions in every case.  As you can see
from the figure, we reproduce well the experimentally observed mass
spectrum of the reaction.  We can see the $\Lambda(1405)$ peak appears
at 1420 MeV in $\pi^+\Sigma^-$ invariant mass instead of nominal 1405 MeV as in
Ref.~\cite{Jido:2009jf}, without contamination from $\Sigma(1385)$ nor
$p$-wave background contribution.  Furthermore, it is important that
$\Sigma(1385)$ contribution appears as a shoulder around 1390 MeV
$\pi^+\Sigma^-$ invariant mass, which may explain with the bump
structure in the empirical data.

Next, let us consider the $\Lambda(1405)$ production cross section from the kaon exchange contribution, which is calculated by the formula,
\begin{equation}
\sigma_{\Lambda ^{\ast}}=3\int^{M_{\rm max}}_{M_{\rm min}}dM_{\pi^+\Sigma^-}\frac{d\sigma_{\rm F.G}}{dM_{\pi^+\Sigma^-}}
\label{eq:25}
\end{equation}
with the range $M_{\rm min}=1400$ MeV and $M_{\rm max}=1440$ MeV.
The factor 3 comes from the branching ratio of $\Lambda(1405)\to\pi^+\Sigma^-$, $33\%$.
The $\Lambda(1405)$ production cross section is consistent with the empirical value~\cite{Braun:1977wd}, as obtained previously in Ref.~\cite{Jido:2009jf}.

\subsection{Production of hyperon resonances -- theoretical studies \label{sec:3B}}
For the understanding of the production of the hyperon resonances in the $K^-d$ reaction, it is important to investigate theoretically the production mechanism of the reaction.
Therefore, we make theoretical studies of the $K^-d$ reaction in this subsection.

\subsubsection{$\pi\Sigma$ channel components\label{sec:3B1}}
\begin{figure}[t]
\begin{center}
\includegraphics[width=14.0cm]{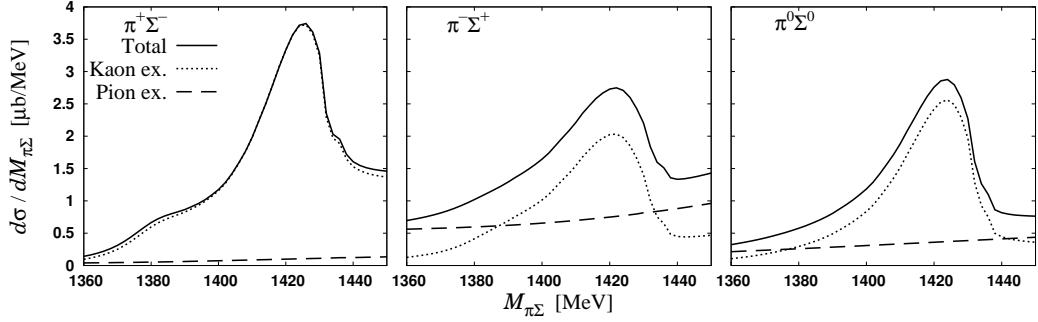}
\caption{\label{fig:3piSigma}
$\pi\Sigma$ invariant-mass spectrum for different $\pi\Sigma$ states at 800 MeV/c incident $K^-$ momentum.
}
\end{center}
\end{figure}

For the clarification of the $\Lambda(1405)$ properties, it is important to understand the behavior of the each $\pi\Sigma$ channel spectrum.
Hence, we plot the $\pi^+\Sigma^-$, $\pi^-\Sigma^+$, and $\pi^0\Sigma^0$ mass spectra in the $K^-d\to(\pi\Sigma)^0n$ reaction in Fig.~\ref{fig:3piSigma}.
Here the initial kaon momentum is fixed as 800 MeV/c.

From Fig.~\ref{fig:3piSigma} one can see that the behavior of the $\pi\Sigma$ mass spectra is slightly different from each other.
This is due to the interference between the $\Lambda(1405)$ contribution ($I=0$) and the $I=1$ non-resonant contribution. The significant non-resonant contribution with $I=1$ around the $\Lambda(1405)$ energy can be seen in the $\pi^-\Lambda$ spectrum shown in Fig.~\ref{fig:piLambda1}.
As a consequence of the interference, the $\pi^+\Sigma^-$ spectrum shows the largest contribution of the $\Lambda(1405)$ production and has the peak position at higher energy than the $\pi^-\Sigma^+$ spectrum, which are 
consistent with the results obtained in Ref.~\cite{Jido:2009jf}.
The interference between the $\Lambda(1405)$ and $I=1$ contributions has been, indeed, important in the photoproduction of 
$\Lambda(1405)$~\cite{Nacher:1998mi,Niiyama:2008rt,Moriya:2009mx}. 
Therefore, experimental data on the $\pi^{\pm} \Sigma^{\mp}$ spectrum will 
bring us further information of the $\Lambda(1405)$ structure.

We also note that the $\pi^0\Sigma^0$ spectrum does not show the $\Sigma(1385)$ contribution, because the $\pi^0\Sigma^0$ channel does not contain $I=1$ component.
Hence, it will be important to observe all the three $(\pi\Sigma)^0$ spectra in the $K^-d$ reaction and compare them in the experiment for the understanding of the $\Lambda(1405)$ structure.

\subsubsection{Diagram contributions\label{sec:3B2}}
\begin{figure}[t]
\begin{center}
\includegraphics[width=14.0cm]{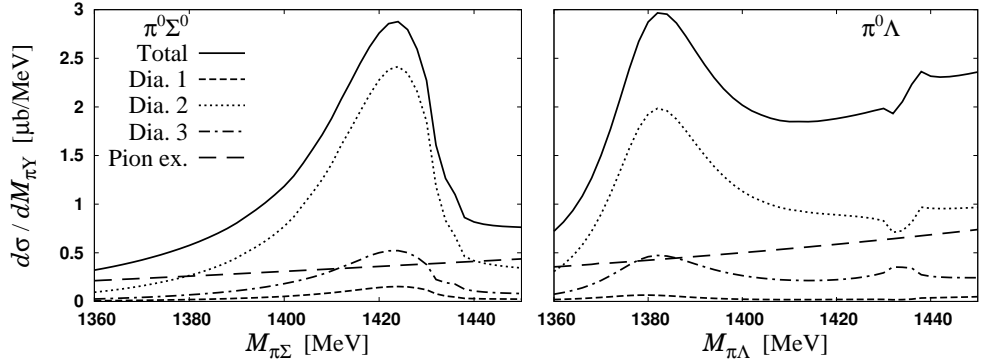}
\caption{\label{fig:pi0}
$\pi^0\Sigma^0$ (left panel) and $\pi^0\Lambda$ (right panel) invariant-mass spectrum separately plotted in each diagram contribution.
The incident $K^-$ momentum is at 800 MeV/c.
The solid line shows the kaon exchange contributions of three diagrams and the pion exchange diagrams.
The dashed , dotted and dash-dotted line show the contribution from diagram 1, 2 and 3 as shown in Fig.~\ref{fig:2}, respectively.
The long-dashed line shows the pion exchange contribution.
}
\end{center}
\end{figure}

For the understanding of the $K^-d$ reaction it is helpful to investigate each diagram contribution to the production of the hyperon resonances.
Here we show the $\pi^0\Sigma^0$ and $\pi^0\Lambda$ invariant mass spectra separately plotted in each diagram contribution in Fig.~\ref{fig:pi0}.
The incident kaon momentum is 800 MeV/c.

As you can see, both  $\Lambda(1405)$ in the $\pi^0\Sigma^0$ spectrum and $\Sigma(1385)$ in the $\pi^0\Lambda$ spectrum show that  diagram 1 (impulse contribution) in the reaction (Fig.~\ref{fig:2}) has quite small contribution, whereas diagram 2 in the reaction (Fig.~\ref{fig:2}) has the largest contribution.
The reason that diagram 1 corresponding to the impulse
production of the hyperon resonances gives small contribution is that deuteron 
hardly has the high momentum component of the inside nucleons as discussed 
in Ref.~\cite{Jido:2009jf}.  
Namely, in order to produce the hyperon
resonances below the ${\bar K}N$ threshold, one needs to create an energetic
nucleon in the final state of the $K^-d\to\pi YN$ reaction.  
Such an energetic nucleon is, however, scarcely produced in the impulse 
process of the $K^-d$
reaction, because large Fermi momentum of the nucleon is highly
suppressed by the deuteron wave function due to the small binding
energy.  As a consequence, the impulse production of the
hyperon resonances has quite small contribution.

Compared with the impulse process, the double scattering process
(diagram 2 and 3) is kinematically favored.  
Namely in the double 
scattering process the transferred energy can be taken from the
incident kaon so that the exchanged kaon has less energy than
on-shell, which is favorable for the production of hyperon resonances
below the ${\bar K}N$ threshold.  
Among the double scattering, since
the $K^-n\to K^-n$ amplitude takes larger value than that of the 
$K^-p\to{\bar K^0}n$
amplitude in the considering energy region, the diagram 2 has the largest contribution to both the
$\Lambda(1405)$ and $\Sigma(1385)$ production (see also discussion in
Ref.~\cite{Jido:2009jf}).

\subsubsection{Angular dependence of production\label{sec:3B3}}
\begin{figure}[t]
\begin{center}
\includegraphics[width=14.0cm]{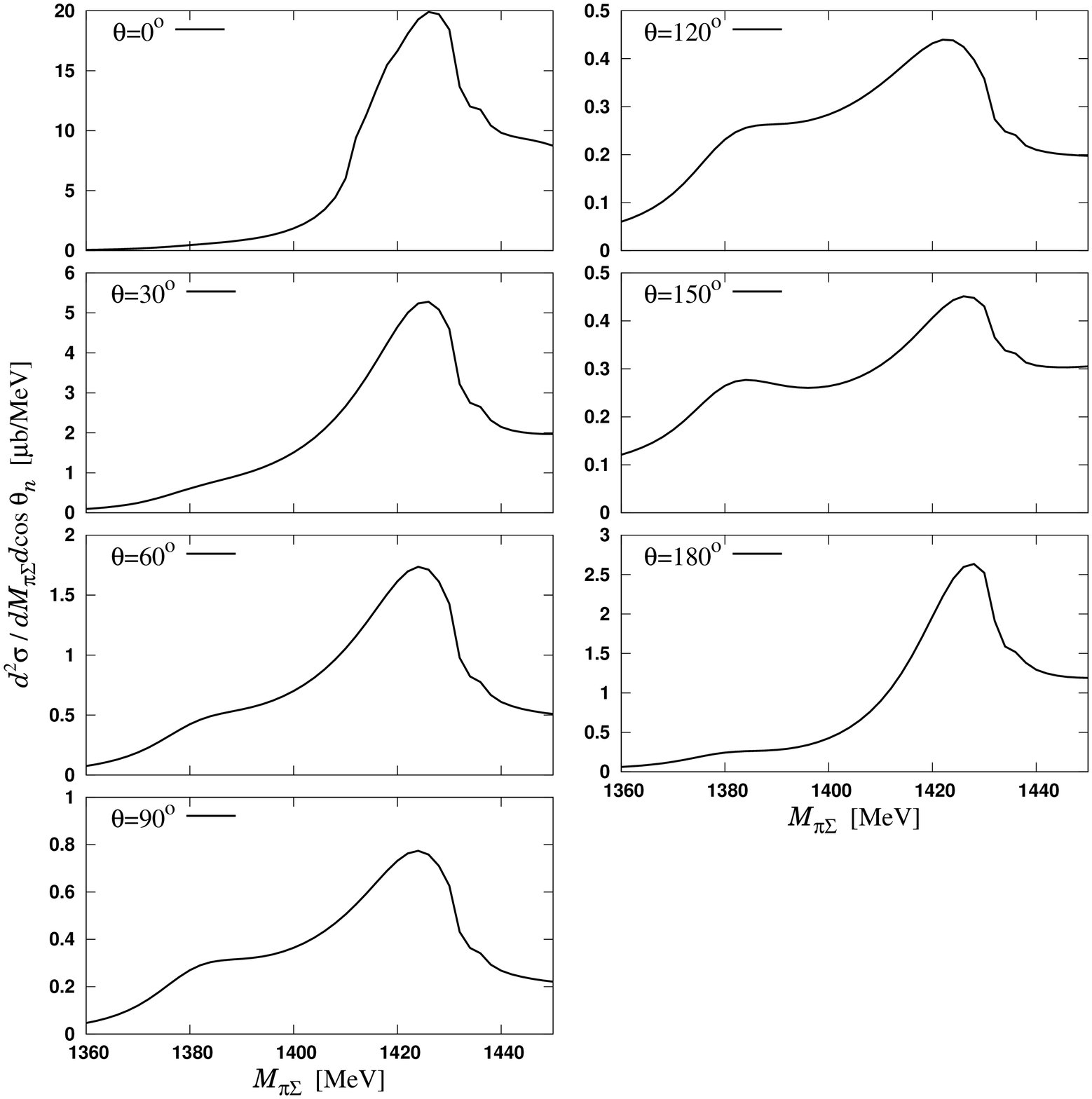}
\caption{\label{fig:angle}
Angular dependence of $\pi^+\Sigma^-$ invariant-mass spectrum at 800 MeV/c incident $K^-$ momentum. The values of $\theta$ in the figure is in unit of degrees. 
}
\end{center}
\end{figure}

Now let us see the angular dependence of the production of the hyperon resonances.
%
%
First of all, we show the double differential cross section $d^2\sigma/dM_{\pi\Sigma}d\cos\theta_n$ in the $K^-d\to\pi^+\Sigma^- n$ reaction in Fig.~\ref{fig:angle}, in which 
the incident kaon momentum is fixed to be 800 MeV/c.
In this reaction we can see the two peaks coming from two resonances, $\Sigma(1385)$ around 1385 MeV and $\Lambda(1405)$ around 1420 MeV, for $60^{\circ} \le \theta_{n} \le 150^{\circ}$.

As one can see, the $\Lambda(1405)$ peak has strong angular $\theta_n$ dependence.
The $\Lambda(1405)$ peak takes its largest value at $\theta_n=0$ (deg.), which  corresponds the forward neutron emission in the total center-of-mass frame, and it becomes about 10 times smaller in the region $\theta_n\geq 60$ (deg.).
The reason is as follows. 
The momentum transfer by the exchanged kaon becomes smaller in the 
double scattering process in the region $\theta_n \leq 60$ (deg.), because in 
this case the incident kaon kicks out neutron in the direction of incident kaon 
and gives most of its momentum simply to the neutron in the first step of the 
double scattering. This small momentum transfer makes the exchanged kaon close 
to be on the mass shell, in which the double scattering amplitude takes larger value. 
Hence, due to the kinematical reasons of the dominant double scattering (discussed in the previous section) and of the suitable momentum transfer, the $\Lambda(1405)$ peak has a large angular dependence and consequently  backward $\Lambda(1405)$ production is dominated.

The $\Sigma(1385)$ peak, on the other hand, shows only small angular dependence in the $K^-d\to\pi^+\Sigma^- n$ reaction. 
This is caused by the $p$-wave nature of $\Sigma(1385)$ in the meson-baryon scattering.
Namely, since the $p$-wave amplitude depends linearly on the transferred momentum, the $\Sigma(1385)$ cannot be much produced with the neutron forward angle, in which the momentum transfer becomes smaller (as discussed above). 
This momentum transfer becomes larger as the angle $\theta_n$ becomes larger, 
since the incident kaon has to kick out neutron in the opposite direction of 
the kaon momentum in the first step of the double scattering. 
Therefore, as a combination of $p$-wave nature of $\Sigma(1385)$ and suitable 
region of the momentum transfer in the process, $\Sigma(1385)$ production 
shows small angular dependence although in the middle $\theta_n$ region the 
production is not favored by the kinematics.

\begin{figure}[t]
\begin{center}
\includegraphics[width=8.0cm]{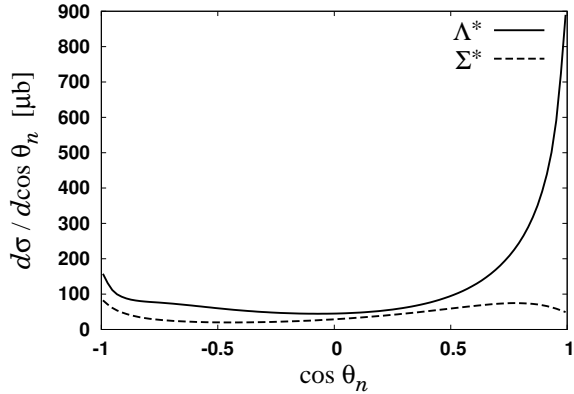}
\caption{\label{fig:angle2}
Angular dependence of the $\Lambda(1405)$ and $\Sigma^0(1385)$ production cross 
sections in $K^{-}d \to \pi ^{0} \Sigma ^{0} n$ (for $\Lambda(1405)$) and 
$\pi ^{0} \Lambda n$ (for $\Sigma^0(1385)$) reactions.
}
\end{center}
\end{figure}

The angular dependence of the hyperon resonance production is clearly seen in the differential cross section $d\sigma/d\cos\theta_n$ with the integration range ($M_{\rm min},~M_{\rm max}$) given in Sec.~\ref{sec:3A}.
In Fig.~\ref{fig:angle2}, we plot the $d\sigma/d\cos\theta_n$ for the $\Lambda(1405)$ ($\Sigma(1385)$) production in the $K^-d\to \pi^0\Sigma^0 n$ ($\pi^0\Lambda n$) reaction.
From Fig.~\ref{fig:angle2}, we can see that the $\Lambda(1405)$ is produced dominantly in the forward neutron angle, whereas the $\Sigma(1385)$ production moderately depends on the neutron angle.

\subsubsection{Missing mass spectrum\label{ref:3B4}}
\begin{figure}[t]
\begin{center}
\includegraphics[width=8.0cm]{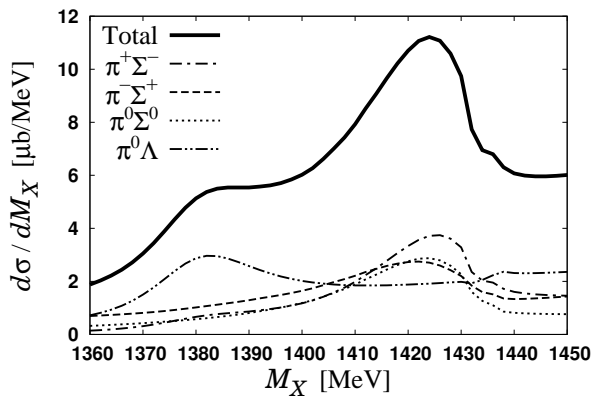}
\caption{\label{fig:missingmass}
Missing mass spectrum in $K^-d\to n X$ process at 800 MeV$/c$ incident $K^-$ 
momentum. The solid, dashed, dashed-dotted, and dashed-double dotted lines 
are total, $\pi ^- \Sigma ^+$, $\pi ^0 \Sigma ^0$, $\pi ^+ \Sigma ^-$, 
and $\pi ^0 \Lambda$ contributions to $X$. 
}
\end{center}
\end{figure}

As we have seen in the previous sections, the $\Lambda(1405)$ peak appears at 1420 MeV instead of nominal 1405 MeV in the $\pi\Sigma$ invariant mass of the $K^-d\to\pi\Sigma n$ reaction thanks to the selective $\Lambda(1405)$ production by the $\bar KN$ channel.
This fact implies that there is a possibility that  the $\Sigma(1385)$ and $\Lambda(1405)$ peaks are seen separately in the $\pi\Sigma$ mass spectrum of the $K^-d$ reaction, which are usually mixed with each other due to the similar peak energies.
Here in order to investigate the behavior of the peak structures, we plot the missing mass spectrum of the reaction $K^-d\to n X$.
In the energy region around 1400 MeV, the missing mass spectrum is expected to be dominated by the $(\pi Y)^0$ system.
Hence, evaluating from the relevant diagrams shown in Figs.~\ref{fig:2} and 
\ref{fig:2.5} and neglecting the amplitudes in which $\pi$ and $Y$ do not 
correlate to each other, which will make no structure in the missing mass 
spectrum, we can plot the missing mass spectrum in Fig.~\ref{fig:missingmass}.
As one can see, there appears the double-peak structure coming from $\Sigma(1385)$ (around 1385 MeV) and $\Lambda(1405)$ (around 1420 MeV). 
In the missing mass spectrum, the $\Sigma(1385)$ peak comes from $\pi^0\Lambda$ and the $\Lambda(1405)$ peak from $(\pi\Sigma)^0$ channel.
The reason why we can separate the $\Sigma(1385)$ peak and the $\Lambda(1405)$ 
one is, as discussed above, we can produce $\Lambda(1405)$ from ${\bar K}N$ 
initial channel, pinning down the production process via conservation of the 
strangeness.


\section{Summary}
\label{sec:4}

We have done the study of the $\Sigma(1385)$ and $\Lambda(1405)$ productions induced 
by $K^-$ on a deuteron target by calculating the $K^-d \to \pi Y N$ reaction.
We have
taken into account both $s$- and $p$-wave contributions of the meson-baryon 
scatterings in the $K^-d$ reaction, especially the contribution of the $\Sigma(1385)$ resonance is newly considered in this study. 
In the $K^-d \to \pi Y N$ reaction, the hyperon resonances are created selectively by the $\bar KN$ channel. Thanks to this fact, the higher pole state of  $\Lambda(1405)$ is produced and the 
$\pi\Sigma$ spectrum has a peak at around 1420 MeV.  This finding implies that 
the $\Lambda(1405)$ and $\Sigma(1385)$ could be seen separately in the missing mass spectrum of the emitted nucleon in the $K^{-} d \to n X$ reaction. 
We have found that invariant mass spectrum of $\pi Y$ is consistent with 
experimental data~\cite{Braun:1977wd}, both for $\Sigma(1385)$ and 
$\Lambda(1405)$ cases. 
We have investigated the angular dependence of the hyperon resonances and found that the $\Lambda(1405)$ production mainly takes place in the backward direction, while the $\Sigma(1385)$ production does not strongly depend on the angle of the emitted nucleon. 
We studied the production mechanisms of $\Sigma(1385)$ and $\Lambda(1405)$ 
from the theoretical side. 
We have also estimated the pion exchange contributions coming from diagrams in which $\pi$ and $Y$ is not correlated to each other and found that these contributions give smooth background and do not spoil the peak structure of the hyperon resonances.

\section*{Acknowledgment}

This work was in part supported by the Grant-in-Aid for Scientific Research
from MEXT and JSPS (Nos.
22740161, 
24105706, %
), 
the collaboration agreement between the JSPS of Japan and the CSIC of Spain,  
and the Grant-in-Aid for the Global COE Program 
``The Next Generation of Physics, Spun from Universality and Emergence''
from MEXT of Japan.
This work is partly supported by DGICYT contract number
FIS2006-03438.  We acknowledge the support of the European Community-Research
Infrastructure Integrating Activity
"Study of Strongly Interacting Matter" (acronym HadronPhysics2, Grant Agreement
n. 227431) under the Seventh Framework Programme of EU.
This work is
part of the Yukawa International Program for Quark-Hadron Sciences
(YIPQS).


%

\vfill\pagebreak

\appendix
\section{Chiral Unitary Model for $\Lambda(1405)$ and $\Sigma(1385)$} 
\label{appendix:A}
Let us briefly formulate the meson-baryon scattering amplitude in
chiral unitary approach.  First of all, in order to include the
higher orbital angular momenta for the meson-baryon system, we
expand the amplitude $T_{ji}(W,x)$ ($i$ and $j$ denote the initial and
final meson-baryon channel, respectively, $W$ the center-of-mass
energy, and $x=\cos\theta$ with the scattering angle $\theta$ in the
center-of-mass frame) into the partial wave as,
\begin{align}
&T_{ji}(W,x)=F_{ji}(W,x)\delta_{s_js_i}-iG_{ji}(W,x)({\hat k}'\times{\hat k})\cdot \mbox{\boldmath $\sigma$}_{s_js_i}~,\\
&F_{ji}(W,x)=\sum^\infty_{l=0}\left[(l+1)f^{(l)}_+(W)+lf^{(l)}_-(W)\right]_{ji}P_l(x)~,\\
&G_{ji}(W,x)=\sum^\infty_{l=1}\left[f^{(l)}_+(W)-f^{(l)}_-(W)\right]_{ji}P'_l(x)~.
\end{align}
Restricting the orbital angular momentum up to $p$-wave ($l=1$), $F$
and $G$ can be written as,
\begin{align}
&F_{ji}(W,x)=f^{(s)}(W)+x(2f^{(p)}_+(W)+f^{(p)}_-(W))~,\\
&G_{ji}(W,x)=f^{(p)}_+(W)-f^{(p)}_-(W)~.
\end{align}
Then, we make unitarizations to the amplitudes in the algebraic 
equation as~\cite{Jido:2002zk},
\begin{align}
&f^{(s)}(W)=(1-f^{(s)}_{\rm tree}g)^{-1}f^{(s)}_{\rm tree}~,\\
&f^{(p)}_+(W)=(1-f^{(p)}_{+{\rm ~tree}}g)^{-1}f^{(p)}_{+~{\rm tree}}~,\\
&f^{(p)}_-(W)=(1-f^{(p)}_{-~{\rm tree}}g)^{-1}f^{(p)}_{-~{\rm tree}}~.
\end{align}
Here $f_{\rm tree}$ corresponds to the tree-level amplitude, giving
the interaction kernel of the coupled-channel.  In this study, we use
the Weinberg-Tomozawa term to $f^{(s)}_{\rm tree}$, same as in
Ref.~\cite{Jido:2009jf},
\begin{equation}
(f^{(s)}_{\rm tree})_{ij} = -C_{ij}\frac{1}{4f^2}(2W-M_i-M_j)\sqrt{\frac{M_i+E_i}{2M_i}}\sqrt{\frac{M_j+E_j}{2M_j}}~.
\end{equation}
with the channel indices $i$ and $j$, the baryon mass $M$, the meson decay constant $f=1.123f_\pi$ ($f_\pi=93$ MeV), and the baryon energy $E$.
For the $f^{(p)}_{\pm~{\rm tree}}$, on the
other hand, we choose the explicit $\Lambda$, $\Sigma$, and $\Sigma^*$
Born terms, same as Eqs.~(19)-(22) in Ref.~\cite{Jido:2002zk}.
The meson-baryon loop integral $g$ in dimensional regularization are written as
\begin{eqnarray}
g_l(W)&=&i2M_l\int\frac{d^4q}{(2\pi)^4}\frac{1}{(P-q)^2-M^2_l+i\epsilon}\frac{1}{q^2-m_l^2+i\epsilon}\nonumber\\
&=&\frac{2M_l}{16\pi^2}\Bigl\{a_i(\mu)+\ln\frac{M^2_l}{\mu^2}+\frac{m_l^2-M_l^2+W^2}{2W^2}\ln\frac{m_l^2}{M_l^2}\nonumber\\
&&+\frac{\bar q_l}{W}[\ln(W^2-(M_l^2-m_l^2)+2{\bar q_l}W)+\ln(W^2+(M_l^2-m_l^2)+2{\bar q_l}W)\nonumber\\
&&-\ln(-W^2+(M_l^2-m_l^2)+2{\bar q}W)-\ln(-W^2-(M_l^2-m_l^2)+2{\bar q_l}W)]\Bigr\}~,\nonumber\\
\end{eqnarray}
where $m$ and $M$ are the meson and baryon masses, respectively, $\mu$ is a regularization scale and $a_i$ are subtraction constants in each of the isospin channels.
Here we use same parameter as Refs.~\cite{Jido:2009jf} and \cite{Jido:2002zk},
\begin{eqnarray}
&&a_{{\bar K}N}=-1.84,~a_{\pi\Sigma}=-2.00,~a_{\pi\Lambda}=-1.83\nonumber\\
&&a_{\eta\Lambda}=-2.25,~a_{\eta\Sigma}=-2.38,~a_{K\Xi}=-2.67~.
\end{eqnarray}

Using the meson-baryon scattering amplitudes in the chiral unitary
approach, we can calculate the cross sections of the ${\bar K}N$ to several
channels.  The cross sections are expressed as,
\begin{equation}
\frac{d\sigma_{ij}}{d\Omega}=\frac{1}{16\pi^2}\frac{M_iM_j}{s}\frac{k'}{k}\left\{|F_{ji}(W,x)|^2+|G_{ji}(W,x)|^2\sin^2\theta\right\}~,
\end{equation}
with initial and final center-of-mass momenta $k$ and $k^{\prime}$.

\section{Meson-baryon scattering amplitudes}
\label{AppB}

In this appendix, we show a way to transform the amplitude obtained in the center of mass frame of the two body meson-baryon system to the amplitude in the baryon rest frame. The idea is that we first obtain the invariant amplitude from the cm amplitude. Next we make the transformation of the invariant amplitude to the baryon rest frame. 

\subsection{Invariant amplitude of meson-baryon scattering}

The Lorentz invariant scattering amplitude of meson-baryon 
can be written in general 
in terms of two Lorentz invariant functions, $A$ and $B$:
\begin{equation}
   T = \bar u(p_{2}) \left[ A(s,t) + \gamma \cdot K B(s,t)\right] u(p_{1})
\end{equation}
where $p_{1}$ and $p_{2}$ are the initial and final baryon momenta, respectively, the four-vector $K$ is defined by
$K=\frac{1}{2}(k_{1}+k_{2})$ with the initial and final meson momenta $k_{1}$
and $k_{2}$, and the Mandelstam variables $s$ and $t$ are given by $s=(p_{1}+k_{1})^{2}$ and $t=(k_{1}-k_{2})^{2}$. 

\subsection{Scattering amplitudes in the c.m.\ frame}

In the c.m.\ frame, the scattering amplitudes are written as
\begin{equation}
   T(W,x) = F(W,x) - i G(W,x) \, (\hat k_{2} \times \hat k_{1}) \times \vec \sigma 
\end{equation}
with the partial wave decomposition 
\begin{eqnarray}
   F(W,x) &=& \sum_{\ell =0}^{\infty} \left[ (\ell +1 ) f_{\ell+}(W) + \ell f_{\ell -}(W) \right]  P_{\ell}(x) \\
   G(W,x) &=& \sum_{\ell =1}^{\infty} \left[  f_{\ell +}(W) - f_{\ell -}(W) \right] P_{\ell} ^{\prime}(x)
\end{eqnarray}
where $W=\sqrt s$ is the c.m.\ energy and $x=\cos\theta$ with  the scattering angle $\theta$ in c.m.\ frame. 
The relation between $(A,B)$ and $(F, G)$ is given by
\begin{eqnarray}
   A &=& \frac{1}{2W} \left[
    \left( W+\frac{M_{1}+M_{2}}{2} \right) \frac{F+xG}{a_{1}a_{2}} 
    +\left( W-\frac{M_{1}+M_{2}}{2} \right) \frac{G}{b_{1}b_{2}} \right] 
    \label{eq:AFG}
    \\
   B &=& \frac{1}{2W} \left[\frac{F+xG}{a_{1}a_{2}} 
  -\frac{G}{b_{1}b_{2}} \right]  \label{eq:BFG}
\end{eqnarray}
where $a=\sqrt{(E+M)/(2M)}$ and $b=\sqrt{(E-M)/(2M)}$ with the c.m.\ baryon 
energy $E$ and baryon mass $M$.

\subsection{Scattering amplitudes in the baryon rest frame}

Let us consider the initial baryon at rest.
The amplitude in this frame is given by
\begin{eqnarray}
 T &=& F^{\prime}(W,x^{\prime}) - iG^{\prime}(W,x^{\prime}) (\hat k_{2}^{\prime} \times \hat k_{1}^{\prime}) \cdot \vec \sigma 
\end{eqnarray}
where  $x^{\prime}$ is the angle between $\vec k_{1}^{\prime}$ and $\vec k_{1}^{\prime}$. 

The baryon rest frame amplitudes, $F^{\prime}$ and $G^{\prime}$, can be written in terms of the Lorentz invariant amplitudes, $A$ and $B$, as
\begin{eqnarray}
  F^{\prime}(W,x^{\prime}) &=& a_{2}^{\prime} \left[A(s,t) + \frac{1}{2} \left(  
       \omega_{1}^{\prime}+\omega_{2}^{\prime}
       +  \frac{|\vec k_{2}^{\prime}|^{2}-|\vec k_{1}^{\prime}|^{2}}{E_{2}^{\prime}+M_{2}}\right) B(s,t) \right] \label{eq:FlabAB}  \\
  G^{\prime}(W,x^{\prime}) &=& - a_{2}^{\prime}  \frac{|\vec k_{1}^{\prime}|\, |\vec k_{2}^{\prime}| }{E_{2}^{\prime}+M_{2}}\, B(s,t)  \label{eq:GlabAB}
\end{eqnarray}
where $a_{2}^{\prime}=\sqrt{(E^{\prime}_{2}+M_{2})/(2M_{2})}$ with the final baryon 
energy $E^{\prime}_{2}$ in the initial baryon rest frame and the final baryon mass $M_{2}$.

The kinematical variables are expressed in terms of $W$ and $\cos \theta^{\prime}$ as
\begin{eqnarray}
   |\vec k_{1}^{\prime}| &=& \frac{W |\vec k_{1}|}{M_{1}} \\
   \omega_{1}^{\prime} &=& \gamma W -M_{1} \\
   \omega_{2}^{\prime} &=& \frac{\omega_{2} + \beta \cos \theta^{\prime}
   \sqrt{\omega_{2}^{2} - \gamma^{2}m_{2}^{2}(1-\beta^{2}\cos^{2}\theta^{\prime})}}{\gamma(1-\beta^{2}\cos^{2}\theta^{\prime})} \\
   |\vec k_{2}^{\prime} | &=&\frac{\beta \cos \theta^{\prime} \omega_{2} + 
   \sqrt{\omega_{2}^{2} - \gamma^{2}m_{2}^{2}(1-\beta^{2}\cos^{2}\theta^{\prime})}}{\gamma(1-\beta^{2}\cos^{2}\theta^{\prime})}
\end{eqnarray}

Substituting Eqs.(\ref{eq:AFG}) and (\ref{eq:BFG}) to Eqs.(\ref{eq:FlabAB}) and (\ref{eq:GlabAB}), we obtain
\begin{eqnarray}
  F^{\prime}(W,x^{\prime}) &=& \frac{a_{2}^{\prime}}{4W} \left[
    \left( 2W+ \omega^{\prime}_{1} + \omega^{\prime}_{2}+ M_{1} +M_{2} + \frac{| \vec k_{2}^{\prime}|^{2}  -|\vec k_{1}^{\prime}|^{2}}{E_{2}^{\prime}+M_{2}}\right) \frac{F(W,x)+xG(W,x)}{a_{1}a_{2}} 
   \right. \nonumber \\ && \ \ \ \ \ \  \  \left.
   +   \left( 2W - \omega^{\prime}_{1} - \omega^{\prime}_{2}- M_{1} -M_{2} - \frac{| \vec k_{2}^{\prime}|^{2}  -|\vec k_{1}^{\prime}|^{2}}{E_{2}^{\prime}+M_{2}}\right) \frac{G(W,x)}{b_{1}b_{2}}\right]  
    \\
  G^{\prime}(W,x^{\prime}) &=& - \frac{a_{2}^{\prime}}{2W}  \frac{|\vec k_{1}^{\prime}|\, |\vec k_{2}^{\prime}| }{E_{2}^{\prime}+M_{2}}\, 
  \left[\frac{F(W,x)+xG(W,x)}{a_{1}a_{2}} 
  -\frac{G(W,x)}{b_{1}b_{2}} \right]
\end{eqnarray}
where $x=\cos\theta$ and $x^{\prime}=\cos \theta^{\prime}$ with the angles between the mesons in the c.m.\
and baryon rest frames, respectively. These are related by the boost transformation:
\begin{eqnarray}
   \cos \theta = \frac{\gamma}{|\vec k_{2}|}
   (-\beta \omega_{2}^{\prime} + | \vec k_{2}^{\prime}| \cos \theta^{\prime} ) =\eta + \xi x^{\prime}
\end{eqnarray}
where $\beta = |\vec p_{1}|/E_{1}$ and $\gamma = (1-\beta^{2})^{-1/2}$.

The amplitudes $F^{\prime}(W,x^{\prime})$ and $G^{\prime}(W,x^{\prime})$ can be decomposed by partial wave in Lab.\ frame.
\begin{eqnarray}
   F^{\prime}(W,x^{\prime}) &=& \sum_{\ell =0}^{\infty} F^{\prime (\ell)} P_{\ell}(x^{\prime}) =\sum_{\ell =0}^{\infty} \left[ (\ell +1 ) f_{\ell+}^{\prime}(W) + \ell f_{\ell -}^{\prime}(W) \right]  P_{\ell}(x^{\prime})
\\
   G^{\prime}(W,x^{\prime}) &=&  \sum_{\ell =1}^{\infty} G^{\prime(\ell)} P_{\ell} ^{\prime}(x^{\prime}) =\sum_{\ell =1}^{\infty} \left[  f_{\ell +}^{\prime}(W) - f_{\ell -}^{\prime}(W) \right] P_{\ell} ^{\prime}(x^{\prime})
\end{eqnarray}
where
\begin{eqnarray}
   F^{\prime (\ell)}(W) &=& \frac{2\ell +1}{2}
   \int_{-1}^{1}dx^{\prime} F^{\prime}(W,x^{\prime}) P_{\ell}(x^{\prime}) \\
   G^{\prime (\ell)}(W) &=& \frac{2\ell +1}{2\ell (\ell+1)}
   \int_{-1}^{1}dx^{\prime} (1-x^{\prime 2})G^{\prime}(W,x^{\prime}) P_{\ell}^{\prime}(x^{\prime})
\end{eqnarray}
Here we have used 
\begin{eqnarray}
   \int_{-1}^{1} dx P_{\ell}(x) P_{k}(x) &=&
   \frac{2}{2\ell +1}\delta_{\ell k} \\
    \int_{-1}^{1} dx (1-x^{2}) P^{\prime}_{\ell}(x) P^{\prime}_{k}(x) &=&
    \int_{-1}^{1} dx  P^{1}_{\ell}(x) P^{1}_{k}(x) =
   \frac{2}{2\ell +1} \frac{(\ell+1)!}{(\ell-1)!} \delta_{\ell k} .
\end{eqnarray}

\end{document}